\documentclass[fleqn,usenatbib]{mnras}

\usepackage{newtxtext,newtxmath}
\usepackage[T1]{fontenc}

\DeclareRobustCommand{\VAN}[3]{#2}
\let\VANthebibliography\thebibliography
\def\thebibliography{\DeclareRobustCommand{\VAN}[3]{##3}\VANthebibliography}

\usepackage{graphicx}	
\usepackage{amsmath}	
\usepackage{caption} 
\usepackage{threeparttable}

\title[Properties of Fibrils]{Properties of Chromospheric Fibrils Around a Quiescent Filament}

\author[Qifan Dong et al.]{
Qifan Dong,$^{1,2}$
Xiaoli Yan,$^{1,3}$\thanks{E-mail: yanxl@ynao.ac.cn}
Zhike Xue,$^{1,3}$
Jincheng Wang,$^{1,3}$
Zhe Xu,$^{1,3}$
Liheng Yang,$^{1,3}$
Yian Zhou,$^{1}$
\newauthor
Xinsheng Zhang,$^{1}$
Zongyin Wu$^{1,2}$
and Guotang Wu$^{1,2}$
\\
$^{1}$Yunnan Observatories, Chinese Academy of Sciences, Kunming 650216, People's Republic of China\\
$^{2}$University of Chinese Academy of Sciences, Beijing 100049, People's Republic of China\\
$^{3}$Yunnan Key Laboratory of Solar Physics and Space Science, Kunming 650216, People's Republic of China
}

\date{Accepted XXX. Received YYY; in original form ZZZ}

\pubyear{\the\year{}}

\begin{document}
\label{firstpage}
\pagerange{\pageref{firstpage}--\pageref{lastpage}}
\maketitle

\begin{abstract}
Fibrils are dynamic plasma structures in the solar chromosphere. Studying these structures is critical for understanding solar atmospheric heating and mass transportation. The purpose of this study is to obtain the characteristics of fibrils surrounding the filament. By employing high-resolution H$\alpha$ data obtained from the New Vacuum Solar Telescope (NVST), we undertake a detailed analysis of the properties of 63 fibrils situated in the vicinity of the filament. Comparing the fibrils on both sides of the filament demonstrates that these fibrils have similar physical properties except for their orientation. The properties of fibrils are statistically measured, including lifetimes of 150–650 s, widths of 320–850 km, maximum lengths of 3–8.5 Mm, projection velocities of 7–29\,km\,s$^{-1}$, and decelerations of 45–474\,m\,s$^{-2}$. The dominant oscillation period of fibrils is predominantly concentrated in the range of 4.8-6.6 minutes (2.5-3.5 mHz). Transverse oscillations are identified in a subset of fibrils, with periodicities of 269–289 s and phase speeds of 13.7–25.8\,km\,s$^{-1}$, indicating the presence of kink-mode magnetohydrodynamic (MHD) waves.
\end{abstract}

\begin{keywords}
methods: data analysis -- methods: observational --  techniques: image processing -- Sun: chromosphere -- Sun: oscillations.
\end{keywords}

%%%%%%%%%%%%%%%%%%%%%%%%%%%%%%%%%%%%%%%%%%%%%%%%%%

%%%%%%%%%%%%%%%%% BODY OF PAPER %%%%%%%%%%%%%%%%%%

\section{Introduction} \label{sec:intro}

The primary activities in the chromosphere are multi-scale jets, among which small-scale jets can be further classified into spicules, fibrils, mottles and rapid blue/red shifted excursions (RBE/RREs) \citep{Tsiropoula+etal+2012, Sterling+2021, Skirvin+etal+2023}. Solar fibrils are magnetically guided plasma structures predominantly aligned along chromospheric magnetic field lines, with their roots anchored in photospheric magnetic flux concentrations. Their elongated morphology directly manifests the geometric configuration of local magnetic topology, where horizontal fibrils observed on the solar disk transform into vertical spicules in limb regions \citep{Jafarzadeh+Rutten+2017}. Additionally, there is another type of fibrils characterized by higher dynamics, shorter length, and a shorter lifespan, referred to as dynamic fibrils (DFs) \citep{Hansteen+etal+2006, DeP+etal+2007, 2013ApJ...776...56R}. Fibrils, type I spicules, and mottles are generally considered to be manifestations of the same features at different locations, exhibiting analogous physical properties in terms of lifetime, velocity, and spatial scale \citep{1995ApJ...450..411S, 2007PASJ...59S.655D, 2012ApJ...759...18P}. In the quiet regions, these features are referred to as mottles \citep{1963AJ.....68R.273B, 1963ApJ...138..648B, 1997A&A...324.1183T, 2007ASPC..368...65D, 2007ApJ...660L.169R}, while in active regions and plage regions, they are known as fibrils \citep{DeP+etal+2007, 2013ApJ...776...56R}, and at the solar limb, they are classified as type I spicules (as categorized by \citealp{2007PASJ...59S.655D}). These fibrillar structures are believed to serve as conduits for mass and energy transfer between the chromosphere and the corona \citep{1982ApJ...255..743A, 2004Natur.430..536D, 2004AA...424..279T, 2006RSPTA.364..383D, 2012A&A...543A...6M, 2019Sci...366..890S}. The transfer of energy occurs through oscillations and waves.

Fibrils are typically observed in the chromospheric H$\alpha$ line core and the Ca\,{\footnotesize II} 8542\,Å line. Recently, some studies have begun to focus on the observational characteristics of fibrils in the Ca\,{\footnotesize II} K and H lines. It has been found that, compared to the commonly observed chromospheric Ca\,{\footnotesize II} K line, fibrils appear narrower. Based on spectral line inversions, the atmospheric temperature within bright fibrils is 100–200 K higher than that of the surrounding environment \citep{Pietarila+etal+2009, Kianfar+etal+2020}. 

The origin of fibrils remains vague. In an early review, \citet{2000SoPh..196...79S} summarized various models, including the pulse model \citep{1982ApJ...257..345H, 1982SoPh...75...99S, 1993ApJ...407..778S}, Alfv\'en wave model \citep{1999A&A...347..696D, 1999ApJ...514..493K, 2017ApJ...848...38I}, and magnetic reconnection model \citep{2016SoPh..291.3207S, 2017ApJ...836...24G}, etc. Using one-dimensional numerical simulations, \citet{2007ApJ...666.1277H} showed that slow-mode magnetoacoustic shocks drive the fibrils. \citet{DeP+etal+2007} contrasted observational data from the Swedish Solar Telescope (SST) with two-dimensional radiative MHD simulations, concluding that fibrils might originate from shock waves generated by photospheric oscillations leaking along magnetic field lines into the chromosphere. Recently, \citet{Rutten+etal+2019} proposed that the origin of H$\alpha$ fibrils is causally linked to magnetic dynamism that drives intermittent heating events (manifested on the disk as rapid blue/red excursions).

The anomalous heating of the solar atmosphere has been a significant issue. A prevailing explanation is the MHD wave dissipation model \citep{1987ApJ...317..514D, 2012Sci...336.1099K}. Consequently, identifying MHD waves and estimating their energy to determine if they are sufficient to heat the corona have become a primary objective. In recent years, extensive observations of transverse oscillations have been made in both spicular and nonspicular structures \citep{2007Sci...318.1574D, 2009A&A...497..525H, 2011Natur.475..477M, 2011ApJ...736L..24O, 2012ApJ...750...51K, 2012NatCo...3.1315M, 2024A&A...687A.249C}. \citet{2009SSRv..149..355Z} reviewed previous studies on spicule oscillations and waves, finding that the oscillation periods typically range at 80-300 s, and suggested that these transverse oscillations might be caused by kink waves or Alfv\'en waves. \citet{2012ApJ...750...51K, 2013ApJ...779...82K} conducted detailed observational studies specifically on the transverse oscillations of mottles, concluding that these oscillations are induced by MHD kink waves. \citet{Jafarzadeh+So+2017ApJS} investigated the transverse oscillations of Ca\,{\footnotesize II} H fibrils in the lower chromosphere and proposed that Alfv\'{e}nic or kink waves are prevalent in these fibrils.

Although significant progress has been made in observational studies of fibrils, early research was primarily focused on fibrils in typical active regions. \citet{Foukal+1971} and \citet{1998SoPh..182..107M} provided a schematic diagram of a filament and its surrounding structures, briefly noting that the fibrils on either side of the filament exhibit different orientations. Although the observational sample in \citet{2003A&A...402..361T} included regions on both sides of the filament, no classification study was conducted on them. In this study, we present a detailed statistical analysis of the properties of fibrils around the filament. Additionally, we analyzed the transverse oscillations of fibrils in this region and estimated the energy carried by waves.

\section{Observations and methods} 
\label{sec:Observations and methods}

\subsection{The Observations}
\label{subsec:Obs}
The observational data were acquired between 03:57 and 04:51 UT on 2023 November 1, located at N27W03, observed by the 1m New Vacuum Solar Telescope (NVST) at the Fuxian Solar Observatory (FSO), Chinese Academy of Sciences (CAS). The field of view (N27W03) located at solar coordinates (X, Y) = (+31$^{\prime\prime}$, +361$^{\prime\prime}$) with the $\mu$ (cosine of the heliocentric angle) value is 0.926. The H$\alpha$ imaging system used by NVST is a tunable Lyot filter with a bandwidth of 0.25\,Å \citep{2014RAA....14..705L}. The dataset includes simultaneous observations in the H$\alpha$ line center, H$\alpha \pm 0.6$\,Å and TiO bands. To reduce atmospheric disturbances and image degradation, the observational images are reconstructed through a two-level processing \citep{Cai+etal+2022}. Initially, frame selection is performed, followed by reconstruction using the speckle masking \citep{1977OptCo..21...55W, 1983ApOpt..22.4028L, 2016NewA...49....8X}. The reconstructed field-of-view (FOV) is 180$^{\prime\prime}$ × 180$^{\prime\prime}$, with a spatial resolution of 0.33$^{\prime\prime}$ and a temporal resolution of $\sim$10\,s or better \citep{2014RAA....14..705L, Yan+etal+2020}. The effective time resolution of the data we used is 44 s. 

Additionally, we utilize the EUV data and line-of-sight (LOS) magnetograms observed by the Solar Dynamics Observatory (SDO) during this period to supplement the analysis of the impact of fibrils on the upper atmosphere and the magnetic field structure of fibrils. The Atmospheric Imaging Assembly (AIA) detector provided high-altitude atmospheric data, with a spatial resolution of 1.5$^{\prime\prime}$ (pixel size 0.6$^{\prime\prime}$) and a temporal resolution of approximately 12 s \citep{2012SoPh..275...17L}. The Helioseismic and Magnetic Imager (HMI) provided LOS magnetic field data, with a spatial resolution of 0.5$^{\prime\prime}$ per pixel and a temporal resolution of approximately 45 s \citep{2012SoPh..275..229S}.

Furthermore, the Chinese H$\alpha$ Solar Explorer (CHASE) provided full-disk H$\alpha$ images, with a spatial resolution of 1.2$^{\prime\prime}$ and a temporal resolution of approximately 60 s \citep{2019RAA....19..165L}.

\subsection{Methods}
\label{subsec:Methods}
For the measurement of lifetime, our sample data were obtained by visually identifying suitable fibrils. The primary basis for determining the lifespan of fibrils is the time interval between the initial and final observations of the tracked fibril. In the initial and final stages of a fibril's evolution, its structure is relatively small, and surrounding fibrils may obscure the observed fibril. This introduces uncertainty in determining its lifetime, affecting the accuracy of the measurement.

The maximum length of the fibril was determined through manual extraction. First, the end position of the fibril was determined by visually following the evolution of the fibril over its life cycle. Subsequently, the maximum length was measured from the initial position of the fibril during its evolution to its terminal point. It should be noted that the unclear upper contour of fibrils introduces subjectivity when determining the position of maximum length.

The width of a fibril was determined by measuring the full width at half maximum (FWHM) of the intensity profile of the H$\alpha$ line core perpendicular to its length direction. Given the varying intensities of individual fibrils and the non-uniform background intensity in the images, we minimized measurement errors by selecting a position along the fibril at 50–70\% of its maximum length during the phase when it had reached its maximum extent. First, a slit was placed perpendicular to the fibril's axis at the selected position to obtain the intensity profile. A Gaussian fit was then applied to this intensity profile, and the FWHM derived from the fit was taken as the fibril's width. Since fibrils are often densely distributed, care was taken to ensure the slit avoided interference from neighboring fibrils. 

The projection velocity of the fibril onto the plane of the sky was obtained from the time-distance diagrams named as xt-plots, where x represents the length along the slit and t represents time. To acquire the xt-plots for the fibrils, a slit was placed along the direction of fibril motion during its lifetime and performed interpolation to obtain the intensity values along the fibril's motion direction at each time step. The motion trajectories of fibrils typically exhibit a parabolic shape (including both ascending and descending phases), which is one of the reasons they are considered the disk counterparts of type I spicules. The steepest positions of the trajectory at each time step were identified. By analyzing these steepest positions, the projection velocity of the fibril was calculated. Since the observational target was not located at the solar disk center, the influence of projection effects on the accuracy of fibril velocity measurements resulted in an underestimation of these values in the derived results.

Similar to the projection velocity, the deceleration was also obtained from the xt-plots. Note that the term "deceleration" refers to the acceleration of the projection velocity. Due to the dense distribution of fibrils, the evolution of an individual fibril in the xt-plots was influenced by adjacent fibrils, making it difficult to fit the trajectory of the fibril based solely on the intensity data from the xt-plots. Therefore, we first extracted the data corresponding to the parabolic trajectory of the fibril, and then performed a parabolic fit on the extracted data to obtain the deceleration.

\begin{figure*}
	\includegraphics[scale=1]{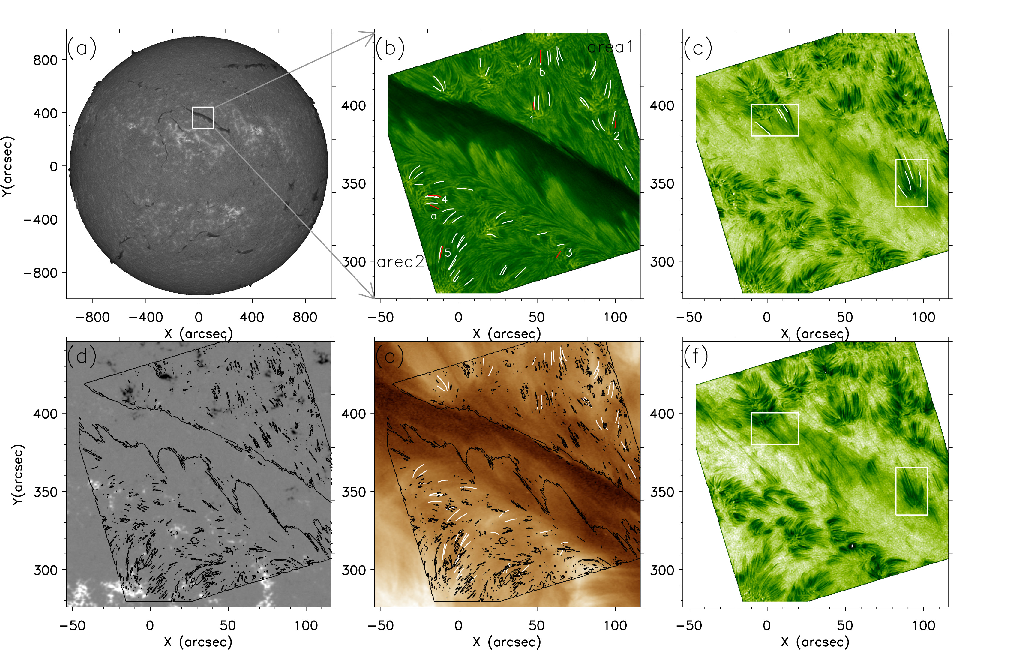}	
	\caption{Appearance of the filament in multiwavelength observations. Panel (a): the full-disk H$\alpha$ image from CHASE, the white box region represents the field of view observed by NVST. Panel (b):  H$\alpha$ line core image with 63 detected fibrils labeled by white and red lines, the upper part of the filament is designated as area 1, and the lower part is designated as area 2. The numbers and red lines indicate the positions of fibrils in Figure \ref{fig:xtplot}. Fibril ``a'' is taken as a sample shown in Figure \ref{fig:Oscillation}. Fibril ``b'' is taken as a sample shown in Figure \ref{fig:Oscillation2}. Panel (c): H$\alpha$ - 0.6\,Å blue wing image. The white box outlines the fibril beneath the filament, with fibrils positions labeled by white lines. Panel (d): the LOS magnetogram from SDO/HMI, the location of the filament is overlaid on the LOS magnetogram. Panel (e): 193\,Å image from SDO/HMI with 63 detected fibrils labeled by white lines.  Panel (f): H$\alpha$ + 0.6\,Å red wing image. The white box outlines the fibril beneath the filament.}	
	\label{fig:FOV}	
\end{figure*}

\section{Result} 
\label{sec:Result}

\subsection{Properties of Chromospheric Fibrils}
\label{subsec:Properties}
Figure \ref{fig:FOV} illustrates co-temporal and co-spatial observational images of the H$\alpha$ line core and wing image, 193\,Å image, and LOS magnetograms in a quiescent filament region. In reference to the filament within the FOV, one region around the filament is designated as area 1, and the other region is designated as area 2 (see Figure \ref{fig:FOV}(b)). Fibrils surrounding filament channels exhibit chirality similar to filament barbs \citep{1998SoPh..182..107M, 2010MmSAI..81..673P}. In this region, due to the influence of the filament channel, fibrils in area 2 generally align in the same direction as the filament barbs, trending southwestward, while those in area 1 trend northeastward. Additionally, the magnetic polarity of the fibril footpoints in area 1 is negative, while in area 2, it is positive (Figure \ref{fig:FOV}(d)). This is because area 1 and area 2 are located on opposite sides of the filament channel, which originates from the magnetic inversion boundary. 

We identified 71 fibrils during the time period of 03:57-04:51 UT on 2023 November 1 from the fibrils around the filament. The criteria for selecting fibrils are as follows: having a complete evolutionary process, being as independent as possible and showing no crossing with surrounding fibrils. Among these, 63 fibrils were identified within the H$\alpha$ line core and were located in the regions on either side of the filament (30 fibrils in area 1 and 33 fibrils in area 2) (see Figure \ref{fig:FOV}(b)), and 8 fibrils were identified within the H$\alpha$ - 0.6\,Å band and were located below the filament (see Figure \ref{fig:FOV}(c)). The reason for using different bands for identifying fibrils is twofold: (1) for the fibrils beneath the filament that cannot be observed in H$\alpha$ line core, we used the H$\alpha$ off-band images; (2) for the fibrils around the filament, we used the H$\alpha$ line core images because fibrils in the H$\alpha$ line core exhibit a complete evolutionary compared with those in H$\alpha$ off-band. Figure \ref{fig:FOV}(e) displays the spatial distribution of 63 fibrils in the 193\,Å images. These fibrils are predominantly located on the frozen magnetic structures in the upper atmosphere, which may be associated with the overlying coronal loops of the filament. 

\begin{figure*}
	\centering  
        \vspace{-1.5cm} % 上移3.5cm	
	\includegraphics[scale=1]{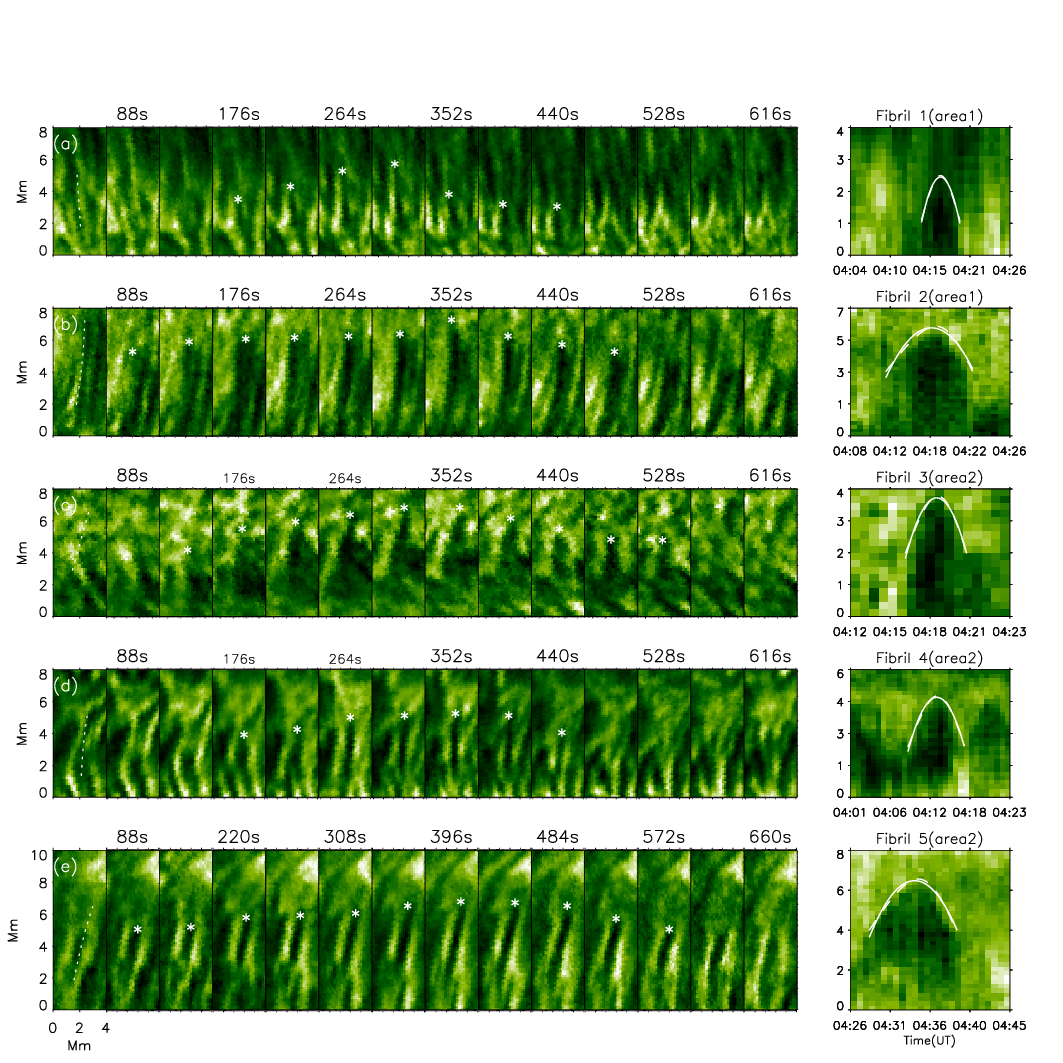}	
	\caption{Space-time plots of fibrils showing the evolution of the fibrils. Panels (a), (b), (c), (d) and (e) correspond to the time series and xt-plots of fibrils 1, 2, 3, 4 and 5 in Figure \ref{fig:FOV}(b) observed in the H$\alpha$ line core. For clarity, the tops of the fibrils are marked with ``*''. In each panel, the dashed line in the first image represents the position and length of the slit. In the xt-plots, the dashed line represents the selected parabolic trajectory, while the solid line indicates the fitted parabolic trajectory.}	
	\label{fig:xtplot}	
\end{figure*}

Figure \ref{fig:xtplot} shows the evolutionary characteristics of the five typical fibrils marked in red in Figure \ref{fig:FOV}(b) at different times, along with their corresponding xt-plots. Panels (a), (b), (c), (d) and (e) of Figure \ref{fig:xtplot} correspond to fibrils 1, 2, 3, 4 and 5 in Figure \ref{fig:FOV}(b), respectively. To facilitate comparison, two samples each were selected from area 1 and area 2, ensuring that the selected fibrils had relatively longer lifetimes. However, the fibril in Figure \ref{fig:xtplot}(a) (fibril 1) has a shorter lifetime, beginning to emerge at 176 s and disappearing by 440 s. Taking Figure \ref{fig:xtplot}(e) as an example, the left side shows the changes in the length of the identified fibril throughout its lifetime, while the right side presents the corresponding xt-plot. The asterisk (*) marks the top of the identified fibril. In the first image of Figure \ref{fig:xtplot}(e), the dashed line indicates the position and length of the slit used to extract data, and the time on the image represents the time interval from the first image. In the xt-plot, two parabolic trajectories are displayed. The dashed line represents the selected parabolic trajectory, while the solid line represents the fitted parabolic trajectory. Our analysis reveals that the fibrils undergo a distinct rise and fall during their lifetimes, with the xt-plots exhibiting clear parabolic trajectories. The motion trajectories of nearly all analyzed fibrils in the xt-plots displayed parabolic shapes, consistent with type I spicules. However, we also identified rapidly evolving fibrils. Due to the temporal resolution limitations of the observational equipment, only the upward motion of these fibrils could be captured, leaving it unclear whether their downward motion was unobserved or whether they were heated during their ascent, rendering them invisible in the H$\alpha$ band.

\begin{figure*}
        \vspace{1.5cm}
	\includegraphics[scale=1.2]{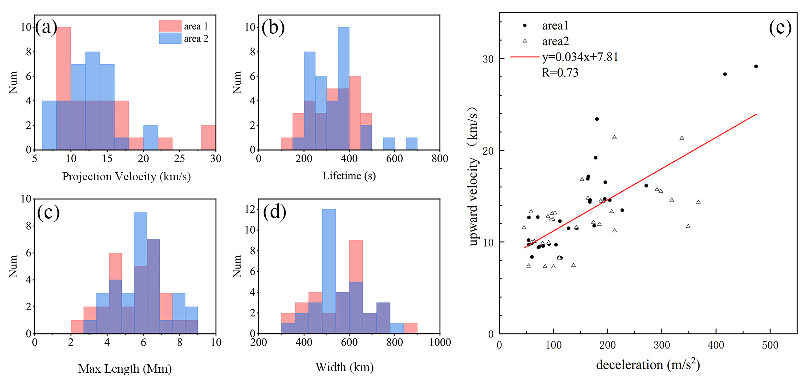}	
	\caption{Panel (a): histogram of projection velocity of the fibrils in area 1 and area 2. Panel (b): histogram of lifetime of the fibrils in area 1 and area 2. Panel (c): histogram of maximum length of the fibrils in area 1 and area 2. Panel (d): histogram of width of the fibrils in area 1 and area 2. Panel (e): Scatter plot of deceleration versus projection velocity for fibrils. The ``$\bullet$'' symbols represent fibrils in area 1, while the ``$\triangle$'' symbols represent fibrils in area 2. The red line represents the result of a linear fit applied to the data, yielding the equation \( y = 0.034x + 7.81 \) with a correlation coefficient of \( R = 0.73 \). }	
	\label{fig:histogram}	
\end{figure*}

Columns 1, 2, and 3 of Table \ref{tab:fibril Properties} present the mean values and standard deviations of projection velocity, lifetime, maximum length, width, and deceleration of fibrils in the regions on either side of the filament, as counted in the H$\alpha$ line core. Figure \ref{fig:histogram}(a)-(d) show the histograms of these properties for areas 1 and 2. 

\begin{table}
        \begin{threeparttable} 
	\centering
	\caption{Properties of fibrils in different areas}
	\label{tab:fibril Properties}
        \setlength{\tabcolsep}{0pt}
	\begin{tabular}{lcccc} 
		\hline
		Region & Area 1 & Area 2 & All\tnote{a} & Area 3\\
		\hline
		{Num} & 30 & 33 & 63 & 8\\
		{$v_{\text{p}}$ (km\,s$^{-1}$)}\tnote{c}  & 13.78 (5.40)\tnote{b} & 12.46 (3.43) & 13.09 (4.49) & 23.60 (5.95)\\
		{T (s)}\tnote{d} & 327 (106) & 324 (112) & 325 (107) & 258 (165)\\
        $L_{\text{max}}$ (Mm)\tnote{e} & 5.50 (1.49) & 5.88 (1.37) & 5.70 (1.46) & 8.09 (2.90)\\
        W (km)\tnote{f} & 568 (132) & 557 (114) & 562 (122) & 498 (75) \\
        a (m\,s$^{-2}$)\tnote{g} & -152.27 (-98.93) & -158.38 (-94.91) & -155.47 (-96.11) & ...\\
		\hline
	\end{tabular}
        \begin{tablenotes}
        \item[a] The mean value of the data from Area 1 and Area 2.
        \item[b] The first data represents the mean value. The values in parentheses represent the standard deviation.
        \item[c] Projection velocity of fibrils.
        \item[d] Lifetime of fibrils.
        \item[e] Maximum length of fibrils.
        \item[f] Width of fibrils.
        \item[g] Deceleration of fibrils.
        \
        \end{tablenotes}
        \end{threeparttable}
\end{table}

The histograms show that the characteristics of area 1 and area 2 are generally similar. The lifetimes of fibrils are evenly distributed between 200 s and 500 s, primarily around 350 s. A few fibrils in area 2 have longer lifetimes, exceeding 10 minutes. The widths are distributed from 350 km to 850 km, with a primary range of approximately 550 km. The maximum lengths are distributed between 3-8.5 Mm, primarily around 6 Mm. The projection velocities in area 1 range from 8 to 29\,km\,s$^{-1}$, while in area 2, they range from 7 to 21\,km\,s$^{-1}$, with the majority averaging between 7 to 18\,km\,s$^{-1}$, except for a few exceeding 20\,km\,s$^{-1}$. The deceleration ranges for area 1 and area 2 are 54 to 474\,m\,s$^{-2}$ and 45 to 367\,m\,s$^{-2}$, respectively. Figure \ref{fig:histogram}(e) shows a scatter plot of deceleration versus projection velocity for fibrils, where ``$\bullet$'' represents fibrils in area 1 and ``$\triangle$'' denotes fibrils in area 2. Consistent with previous studies, a linear relationship is observed between deceleration and projection velocity \citep{2007ApJ...660L.169R, 2012ApJ...759...18P}. The red line in the figure represents the linear fitting result $y$ (km\,s$^{-1}$) = $0.034x$ (m\,s$^{-2}$) + 7.81, with a correlation coefficient of 0.73, indicating a reliable linear relationship between velocity and deceleration. In addition, no significant correlations were found between other parameters in the scatter plots, and the correlation coefficients from the linear fits were all less than 0.5. From the statistical data in Table \ref{tab:fibril Properties}, there are no significant differences in projection velocity, lifetime, maximum length, deceleration and width between fibrils in area 1 and area 2.

Columns 4 and 5 of Table \ref{tab:fibril Properties} present the characteristic data for fibrils in the H$\alpha$ line core including all fibrils in area 1, area 2 (total for area 1 and area 2), and area 3. It should be noted that due to the scarcity of fibrils beneath the filament, statistics could only be conducted in two distinct regions (see Figure \ref{fig:FOV}(c)), resulting in a smaller sample. Nevertheless, from the limited data comparison, it can be found that fibrils beneath the filament have higher velocities, almost double value as those on both sides of the filament, shorter lifetimes, longer maximum lengths, and narrower widths. The discrepancies observed may be attributed to the fact that the line core and line wing reflect different layers of the chromosphere, and the regions inside and outside the filament channel might also differ, as well as the limited sample size.

\subsection{Transverse Oscillations of Chromospheric Fibrils}
\label{subsec:Oscillations}
Due to the limitations of temporal resolution and the complex image structure in the H$\alpha$ line core, the majority of observed fibrils exhibited extremely weak transverse oscillations, making them unanalyzable. To distinguish transverse oscillations from random motions in fibrils, we focused our analysis on three of the most typical examples.

Given the expected sub-pixel oscillation amplitudes, we preprocessed the data using the motion magnification algorithm \citep{Anfinogentov+2016} to ensure measurement accuracy. This technique linearly amplifies displacement amplitudes while preserving oscillation periods unchanged. Although originally developed for SDO/AIA coronal loops, its applicability extends to any time-series imaging data. Validation studies \citep{Zhong+etal+2021, Gao+etal+2022} confirm the algorithm's robustness: (1) strong noise immunity, (2) period preservation, (3) <20\% amplitude error, and (4) reliable detection down to 0.01-pixel amplitudes.

\begin{figure}
	\centering	
        \hspace*{-25pt}
	\includegraphics[scale=0.55]{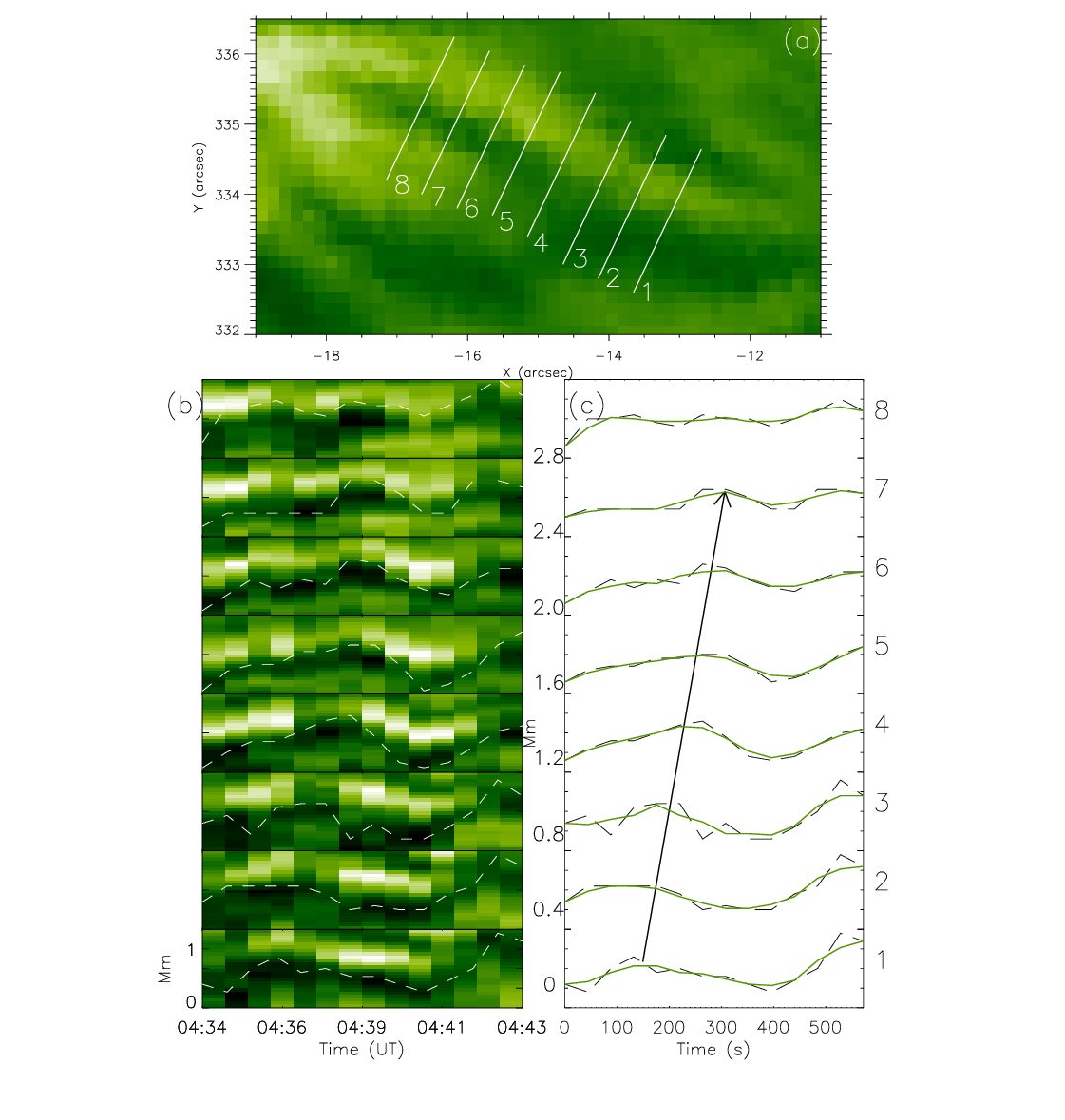}	
        \vspace{-30pt}
	\caption{Panel (a): Motion-magnified view of the fibril labeled ``a'' in Figure \ref{fig:FOV}. The white lines indicate the positions for analyzing transverse oscillations, numbered from 1 to 8 from right to left, with an interval of 0.5 arcsecond. Panel (b): space-time plots corresponding to positions 1-8 in panel (a), displayed from bottom to top. The white dashed lines represent the fitted minimum intensity values for each column. Panel (c): the black dashed lines are the same as the white dashed lines in panel (b). The green line is a smoothed version of the black dashed lines. Black arrows indicate the propagation of waves from right to left in the fibril of panel (a).}	
	\label{fig:Oscillation}	
\end{figure}

\begin{figure*}
	\centering	
	\includegraphics[scale=0.85]{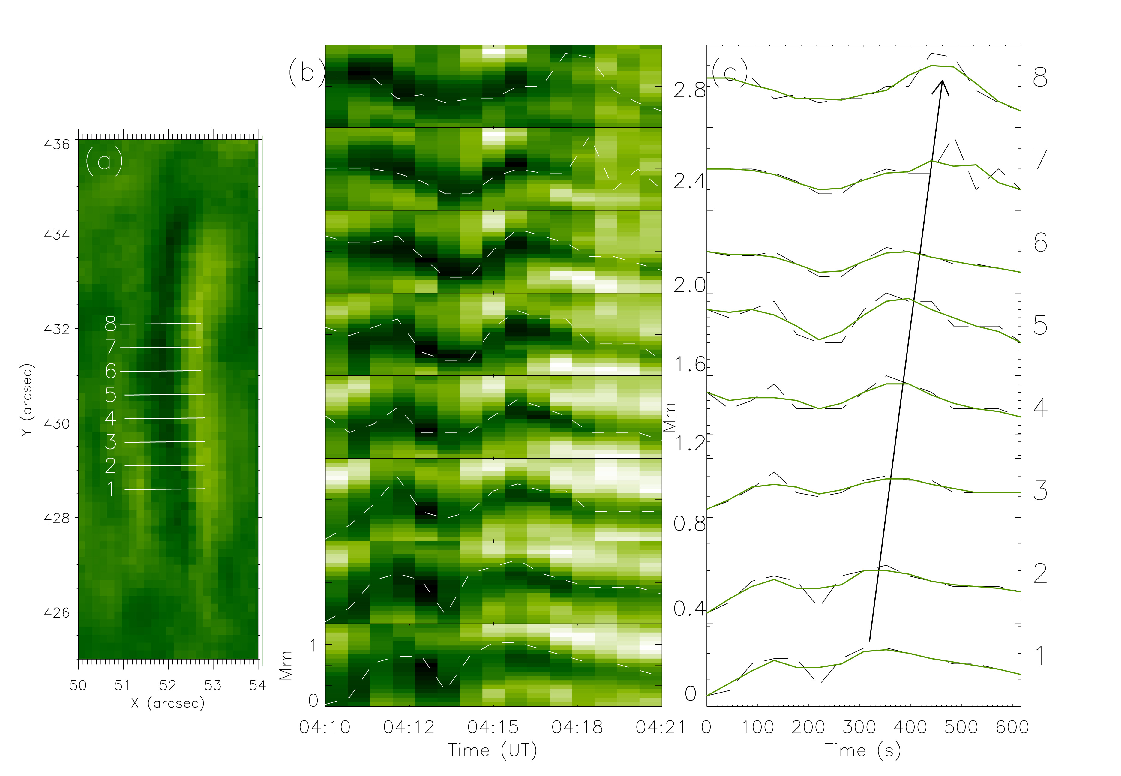}	
	\caption{Same as Figure \ref{fig:Oscillation}. Motion-magnified view of the fibril labeled ``b'' in Figure \ref{fig:FOV}. The black arrow indicate the propagation of waves from bottom to top in the fibril of panel (a).}
	\label{fig:Oscillation2}	
\end{figure*}

Figure \ref{fig:Oscillation} shows the fibril labeled ``a'' in Figure \ref{fig:FOV}(b) and the analysis of its transverse oscillations. To investigate the transverse oscillations, we performed a slicing analysis of the fibril. Figure \ref{fig:Oscillation}(a) displays the motion-magnified fibril structure processed using the algorithm. The white lines indicate the positions for analyzing transverse oscillations, with a spacing of 0.5$^{\prime\prime}$, numbered from 1 to 8. Figure \ref{fig:Oscillation}(b) from bottom to top displays the two-dimensional space-time plots for positions 1-8. The center of displacement for the fibril was determined by selecting the minimum cross-sectional flux at each time frame, which has been marked in the space-time plots. Subsequently, the curve was smoothed over three points to obtain the green line in the Figure \ref{fig:Oscillation}(c). Additionally, the black arrow indicates the wave propagation from right to left along the fibril in panel (a). The oscillation period at each position was determined by Fourier analysis, with an average period of 286 s. The phase velocity of the wave was calculated from the time delay of the maximum displacements at different positions of the fibril using the following formula: $v_{\text{ph}} = L_{\text{max}} / t_L$, where $L_{\text{max}}$ represents the distance between two different positions where the maximum displacement occurs, and $t_{L}$ represents the time difference between the moments when these two positions are reached. The overall phase velocity for positions 1-8 of the fibril was estimated, resulting in an average value of 13.65\,km\,s$^{-1}$. Figure \ref{fig:Oscillation2} presents another example (subregion ``b'' from Figure \ref{fig:FOV}(b)), processed using the same method as Figure \ref{fig:Oscillation}. The black arrow in panel (c) indicates upward wave propagation along the fibril shown in panel (a). This fibril exhibits an average oscillation period of 269 s and phase speed of 16.50\,km\,s$^{-1}$. The phase velocities for the three sample cases ranged from 13.7 to 25.8\,km\,s$^{-1}$, with periods ranging from 269 to 289 s. This range of periods is significantly larger than those reported in previous studies \citep{2009SSRv..149..355Z, 2013ApJ...779...82K, Bate+etal+2022}. The reason for such long periods is that the temporal resolution of the observational equipment is relatively large, allowing only the observation of low-frequency waves. This transverse oscillation can be interpreted as the propagation of waves in space. Given their higher density relative to the ambient atmosphere, fibrils can be modeled as magnetic flux tubes. In contrast to torsional Alfv\'en waves, which typically produce twisting motions without net displacement of the tube axis, the observed transverse displacements are attributed to MHD kink waves propagating along these fibrils formed on the field lines \citep{2009SSRv..149..355Z, 2012ApJ...750...51K, 2013ApJ...779...82K}.

The velocity amplitude of the waves was further calculated. The velocity amplitude can be determined using the formula: $v_A = 2\pi A / P$, where $A$ is the transverse displacement amplitude and $P$ is the period of the wave. The transverse displacement amplitude is the difference between the maximum transverse displacement position and the center position of the fibril, with the center position determined by time-averaging the positions over the entire lifetime of the fibril. The transverse displacement amplitude for the sample cases was 147 $\pm$ 41 km, and the wave period was 286 s, resulting in a velocity amplitude of 3.23 $\pm$ 0.90\,km\,s$^{-1}$. Using the above data, we can estimate the energy flux using the formula: $E = f \rho v_A^2 v_{\text{ph}}$, where $f$ is the density filling factor, $\rho$ is the density of the fibril, $v_A$ is the velocity amplitude, and $v_{\text{ph}}$ is the phase velocity. Here, we used a density filling factor of 5\% as an upper limit \citep{2012NatCo...3.1315M}, and the density of the fibril was 2.2 $\times 10^{-10}$\,kg\,m$^{-3}$ \citep{2004AA...424..279T}. The energy flux was calculated as 0.4 to 6.5\,W\,m$^{-2}$, estimating that the energy of these waves is insufficient to sustain the chromosphere (with a required energy flux of 10$^3$\,W\,m$^{-2}$).

\begin{figure*}
	\includegraphics[scale=1.1]{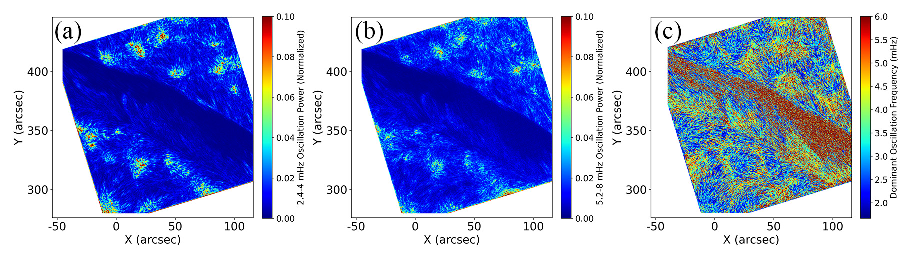}
        %\vspace{-30pt}
	\caption{Panel (a): 2.4-4 mHz power spectrum maps. The oscillation power gradually increases from blue to red in color. The oscillation power of fibrils is significantly higher than that of the surrounding environment, and it reaches the maximum at their footpoints. Panel (b): 5.2-8 mHz power spectrum maps. Panel (c): Dominant oscillation frequency maps. The main oscillation frequency gradually increases from blue to red in color. The frequency at the location of the fibrils is significantly lower than that of the surrounding environment, and it is mainly distributed between 2.5-3.5 mHz, which corresponds to a period of 4.8-6.6 minutes. }	
	\label{fig:fft}	
\end{figure*}

Power spectrum maps and dominant oscillation frequency maps were obtained by performing Fast Fourier Transform (FFT) analysis on the observational dataset (see Figure \ref{fig:fft}). Solar three-minute (5.56 mHz) and five-minute (3.33 mHz) oscillations, which are the key targets of interest, determined the 2.4–8 mHz (2.1-6.9 min) frequency range selected for the computation and visualization of the power spectra in the Figure \ref{fig:fft}(a-b). A blue-to-red color gradient is used to represent the increasing intensity of oscillation power. The analysis shows that fibrils display significantly stronger oscillation power than the surrounding regions, with the maximum power concentrated at their footpoints. This finding confirms the presence of pronounced oscillatory behavior within fibrils. Additionally, a comparison of Figure \ref{fig:fft}(a) and (b) reveals that the oscillation power in the 2.4–4 mHz range is significantly greater than that in the 5.2–8 mHz range, indicating that the oscillation period of the fibril is primarily concentrated in 5 minutes. To further analyze the dominant oscillation frequencies, 15th-order polynomials were used to fit the temporal sequences of each pixel, aiming to reduce boundary effects and spectral leakage. In the dominant frequency map (Figure \ref{fig:fft}(c)), the blue-to-red color gradient corresponds to frequency values ranging from 1.667 to 6 mHz (2.7-10 min). By examining the spatial correlation with the high-power regions in the power spectrum map, it is revealed that the dominant oscillation frequency of fibrils is concentrated in the range of 2.5–3.5 mHz (4.8-6.6 minutes).

\section{Discussion and conclusion} 
\label{sec:conclusion}
While previous investigations of solar fibrils have predominantly focused on active regions and plage areas, the present study conducts a systematic investigation into the physical and oscillatory properties of fibrils surrounding quiescent filaments. Leveraging the high spatial resolution observational capabilities of the NVST, we performed detailed imaging analyses of these fibrils around the filament. It is important to note that the average lifetime of fibrils ranges from 3 to 15 minutes, which can be well resolved by the time resolution of NVST. This allows us to reliably capture their full evolutionary cycle and determine their lifetimes.

\begin{table*}
        \begin{threeparttable}
	\centering
	\caption{Comparison of the properties of fiber-like structures in different works}
	\label{tab:Comparison Properties}
        \setlength{\tabcolsep}{11pt}
	\begin{tabular}{lccccccc} 
		\hline
		 & Type & Band & Lifetime & Length & Width & Velocity & Deceleration \\
          &  &  & (min) & (Mm) & (Mm) & (km\,s$^{-1}$) & (m\,s$^{-2}$) \\
		\hline
		{\citet{2004AA...424..279T}}\tnote{a} & Mottles & H$\alpha$ & 5 & 10 & 1 & 20-40 & ... \\ 
        {\citet{DeP+etal+2007}}\tnote{a} & DFs & H$\alpha$ & 2--11 & 0.4-5.2 & 0.12-0.38 & 8-35 & 120-280 \\
        {\citet{Pietarila+etal+2009}}\tnote{a} & Fibrils & Ca\,\textsubscript{II}\,K & 3-5 & 0.7-7.2 & 0.08-0.13 & 20 & ... \\
        {\citet{2012ApJ...759...18P}}\tnote{b} & Spicules & Ca\,\textsubscript{II}\,H & 3.0-5.7 & 5.3-8.5 & 0.27-0.43 & 22-39 & 182-343 \\
        {This work} & Fibrils & H$\alpha$ & 2.5--10.8 & 3--8.5 & 0.32--0.85 & 7--29 & 46-474 \\
		\hline
	\end{tabular}
        \begin{tablenotes}
        \item[a] On-disk observations.
        \item[b] Off-limb observations. The data are from the active region, where type I spicules dominate.
        \end{tablenotes}
        \end{threeparttable}
\end{table*}

In Table \ref{tab:Comparison Properties}, we compare our statistical results with those of \citet{2004AA...424..279T}, \citet{DeP+etal+2007}, \citet{Pietarila+etal+2009}, and \citet{2012ApJ...759...18P}. Table 2 presents a systematic comparison of dynamic properties across various chromospheric structures, including mottles, DFs, fibrils, and spicules, observed in H$\alpha$ and Ca\,{\footnotesize II} spectral lines. The analysis synthesizes data from multiple observational studies, highlighting key differences in lifetimes, spatial scales, velocities, and deceleration characteristics. As discussed in Section \ref{subsec:Properties}, the fibrils on either side of the filament show no significant property variations except for the horizontal velocities in the photosphere and chromosphere. Therefore, we treat the fibrils within the entire FOV as a single dataset for comparison with previous studies. From the data, DFs and fibrils (this work), both observed in H$\alpha$, show overlapping lifetimes and velocity ranges, though this work reports broader fibril widths and longer lengths compared to DFs. This is consistent with the characteristics of DFs, which are narrower and shorter compared to fibrils \citep{2006RSPTA.364..383D}.  However, the mottles observed in the H$\alpha$ band, in contrast to the DFs and fibrils (in this work), are wider, longer, and have a greater velocity. This might be due to the limitations of the resolution of the early observational equipment. Another reason may be that the statistics of DFs and fibrils (this work) were conducted in the H$\alpha$ line core, whereas \citet{2004AA...424..279T}'s statistics were based on data of the H$\alpha \pm 0.48$\,Å band. We found that the few off-band velocities we collected were similar to those of \citet{2004AA...424..279T}. Comparing the properties of our fibrils with those of type I spicules from \citet{2012ApJ...759...18P}, fibrils exhibit a larger upper limit in lifetime, similar lengths, wider widths, overall slower velocities, and decelerations that are generally comparable except for a few individual cases. There are two possible reasons for these differences: (1) the difference in observational wavelengths, as \citet{Pereira+etal+2016} found that spicules observed in the Ca\,{\footnotesize II} H and H$\alpha$ bands exhibit slight differences in properties; (2) the difference in observational locations, with type I spicules located at the solar limb and fibrils on the solar disk. Projection effects may contribute to the differences in properties. Comparative analysis between the fibrils studied in this work and those observed in Ca\,{\footnotesize II} K lines reveals minimal differences in physical properties apart from width characteristics. This finding is consistent with the results reported by \citet{Pietarila+etal+2009}, where Ca\,{\footnotesize II} K fibrils were likewise found to exhibit narrower widths.

\begin{table*}
        \begin{threeparttable}
	\centering
	\caption{Comparison of the properties of transverse oscillations in different works}
	\label{tab:Comparison Oscillations}
        \setlength{\tabcolsep}{6pt}
	\begin{tabular}{lccccccc} 
		\hline
         & Area & Type & Band & Displacement & Period & Velocity amplitude & Phase speed \\
         &  &  &  & (km) & (s) & (km\,s$^{-1}$) & (km\,s$^{-1}$) \\
		\hline
        {\citet{2012ApJ...744L...5J}} & QS\tnote{a} & On-disk Type I & H$\alpha$; Ca\,\textsubscript{II}\,K & 160-670 & 65-220 & 7-28 & ... \\
        {\citet{2012ApJ...750...51K}} & QS & Mottles & H$\alpha$ & 100-400 & 70-280 & 3-18 & 40-110 \\
        {\citet{Morton+etal+2013ApJ}} & QS & Fibrils & H$\alpha$ & 0-400 & 10-500 & 1-15 & ...  \\ 
        {\citet{Jafarzadeh+So+2017ApJS}} & AR\tnote{b} & Fibrils & Ca\,\textsubscript{II}\,H & 1-91 & 16-199 & 1-4.8 & 1-70\\
        {\citet{2021RSPTA.37900183M}}\tnote{c} & AR & Super-penumbral fibrils & Ca\,\textsubscript{II}\,8542\,Å & 74 & 754 & 0.76 & 25 \\
        {This work} & QS & Fibrils & H$\alpha$ & 95-220 & 269-289 & 1.7-4.8 & 13.7-25.8  \\
		\hline
	\end{tabular}
        \begin{tablenotes}
        \item[a] Quiet sun.
        \item[b] Active region.
        \item[c] The data are average values, while the other authors have given data ranges.
        \end{tablenotes}
        \end{threeparttable}
\end{table*}

Table \ref{tab:Comparison Oscillations} synthesizes key parameters of transverse oscillations across diverse solar features, comparing the results of this study with those of \citet{2012ApJ...744L...5J}, \citet{2012ApJ...750...51K}, \citet{Morton+etal+2013ApJ}, \citet{Jafarzadeh+So+2017ApJS}, and \citet{2021RSPTA.37900183M}, and revealing distinct trends between quiet Sun (QS) and active region (AR) environments. QS studies generally exhibit larger displacements and broader velocity amplitudes compared to AR fibrils. Periods show significant variation, with QS features spanning 65–500 s versus AR oscillations predominantly below 200 s, except for super-penumbral fibrils at 754 s. The phase velocity of mottles is greater than that of fibrils, regardless of whether they are in the QS or AR. Notably, the velocity amplitudes of fibrils identified in this study are comparable to those of AR fibrils (albeit in different observational passbands), with overall values lying between typical measurements for AR and QS fibrils. We speculate that this phenomenon results from the magnetic structure surrounding the filament channel. Of course, time resolution and sample size may also have some influence. Observed discrepancies likely arise from differences in spatial resolution, spectral bandpasses (H$\alpha$ vs. Ca\,{\footnotesize II}), and substructure morphologies (fibrils vs. mottles), emphasizing the sensitivity of oscillation properties to observational techniques and target feature characteristics.

\begin{figure*}
        \vspace{-10pt}
	\includegraphics[scale=1.2]{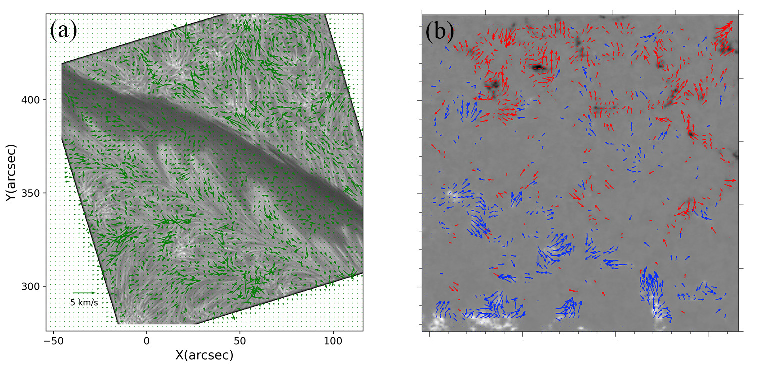}	
	\caption{Panel (a): The 54 minute time-averaged velocity field. Green arrows represent the direction and relative magnitude of fibril velocities. Panel (b): Two-dimensional transverse velocity field derived from the HMI magnetic field, with blue arrows indicating the velocity of positive polarity and red arrows indicating the velocity of negative polarity.}	
	\label{fig:velo}	
\end{figure*}

To investigate the relationship between chromospheric fibrils and their rooted photospheric magnetic fields, we calculated the chromospheric velocity field and the transverse velocity field derived from the HMI magnetic field. To investigate the collective velocity orientation of fibrils, Fourier-based local correlation tracking (FLCT; \cite{Fisher+2008}) was implemented to derive the two-dimensional velocity field across the FOV, as illustrated in Figure \ref{fig:velo}(a). The tracking parameters were configured with: a temporal cadence of 44 s (equivalent to one time resolution), pixel scale of $0.165 \times 725\,\mathrm{km}$ per edge, and Gaussian smoothing kernel $\sigma = 10$. A total of 73 consecutive measurements obtained throughout the observational timeframe were temporally averaged to statistically determine the predominant fibril motion patterns. The velocity vectors in Figure \ref{fig:velo}(a) are represented by green arrows, where directional orientation denotes motion azimuth and arrow length scales with the relative velocity magnitude. The analysis reveals that there is a distinct directionality of the fibrils on both sides of the filament: fibrils in area 1 demonstrate systematic northeastward displacement, whereas those in area 2 exhibit a southwestward motion opposite to that of area 1, aligning with the barb orientation. The two-dimensional transverse velocity field derived from the HMI magnetic field within the time range from 4:00 to 5:00 (Figure \ref{fig:velo}(b)) exhibits the same directionality as the chromospheric velocity field, with blue arrows indicating the direction of motion for the positive polarity magnetic field and red arrows indicating that for the negative polarity magnetic field. The correspondence between the transverse velocity field derived from the HMI magnetic field and the chromospheric velocity field may suggest that the direction of chromospheric fibrils is modulated by the transverse velocities of photospheric magnetic fields.

In summary, we utilized observational data provided by NVST to carry out a statistical study of fibrils around the filament. The fibrils around the quiescent filament had lifetimes ranging from 150-650 s, maximum lengths of 3-8.5 Mm, widths of 320-850 km, projection velocities of 7-29\,km\,s$^{-1}$, and decelerations of 45–474\,m\,s$^{-2}$. The fibrils observed in the H$\alpha$ line core in this work exhibit longer maximum lifetimes (compared to Ca\,{\footnotesize II} K fibrils and spicules), larger spatial scales (wider widths and longer lengths than DFs and Ca\,{\footnotesize II} K fibrils), slower overall velocities (relative to type I spicules), and similar lengths and deceleration characteristics to limb spicules. Notably, their width is the significant difference compared to Ca\,{\footnotesize II} K fibrils, aligning with previous findings of narrower Ca\,{\footnotesize II} K structures. The difference in the directionality of fibril motions on either side of the filament may be related to the motion direction of the underlying photospheric magnetic field.

We found the presence of low-frequency waves in the fibrils on both sides of the filament, with phase velocities ranging from 13.7 to 25.8\,km\,s$^{-1}$, periods of 269 to 289 s, and velocity amplitudes of 1.71 to 4.79\,km\,s$^{-1}$. Using these data, we estimated that the energy of the wave is insufficient to sustain the chromosphere (with a required energy flux of 10$^3$\,W\,m$^{-2}$). Additionally, the existence of the five minute oscillation period and transverse oscillations in fibrils supports the view that the fibrils are driven by magnetoacoustic shocks \citep{DeP+etal+2007}. 

Fibrils are ubiquitous in the solar chromosphere and carry MHD waves, suggesting that these fibrils may contribute to the energy supply of the corona. Additionally, \citet{1998SoPh..182..107M} and \citet{2010MmSAI..81..673P} provided sketches illustrating that fibrils on both sides of filament have different directions, attributing this phenomenon to the influence of the filament channel, but without detailed explanations. Therefore, a more comprehensive and systematic observational analysis is required to investigate the potential relationship between fibrils and filaments.

\section*{Acknowledgements}

This work is sponsored by the Strategic Priority Research Program of the Chinese Academy of Sciences, Grant No. XDB0560000, the National Natural Science Foundation of China (NSFC)  under the numbers Nos. 12325303, 11973084, 12003064, and 12203097, Yunnan Key Laboratory of Solar Physics and Space Science under the number No. 202205AG070009, and Yunnan Fundamental Research Projects under the numbers Nos. 202301AT070347 and 202301AT070349.

%%%%%%%%%%%%%%%%%%%%%%%%%%%%%%%%%%%%%%%%%%%%%%%%%%
\section*{Data Availability}

Upon reasonable request, the data underlying this article will be shared by the corresponding author.


\bibliographystyle{mnras}
\bibliography{references}

%%%%%%%%%%%%%%%%%%%%%%%%%%%%%%%%%%%%%%%%%%%%%%%%%%

% Don't change these lines
\bsp	% typesetting comment
\label{lastpage}
\end{document}